\documentstyle[twocolumn,aps,epsf]{revtex}
\begin{document}
\title{Operation of Quantum Cellular Automaton 
cells with more than two electrons}
\author {M. Girlanda, M. Governale, M. Macucci, G. Iannaccone}
\address{Dipartimento di Ingegneria dell'Informazione,
Universit\`a di Pisa\\
Via Diotisalvi, 2, I-56126 Pisa, Italy}
\date{\today}
\maketitle
\begin{abstract}
We present evidence that operation of QCA (Quantum Cellular Automaton) 
cells with four dots is possible with an occupancy of $4N+2$ electrons 
per cell ($N$ being an integer). We 
show that interaction between cells can be described in terms of a revised
formula for cell polarization, which is based only on the difference between
diagonal occupancies.
We validate our conjectures with full quantum 
simulations of QCA cells for a number of electrons varying from 2 to 6, 
using the Configuration-Interaction method. 
\end{abstract}
\narrowtext
\section{Introduction}
The concept of logic circuits based on Quantum Cellular Automata (QCA), first
proposed by Lent {\sl et al.}\cite{LentAPL}, has received much attention
in the last few years, due to the perspectives of extremely low power 
operation and to the drastic reduction of interconnections it would allow.
The basic QCA building block is represented by a bistable cell made up
of four quantum dots or metallic islands at the vertices of a square
and containing two electrons that can align along the two different 
diagonals, thus encoding the two logical states. For an isolated cell,
alignment along either diagonal is equally likely, but, in the presence
of an external electric field such as that due to a nearby cell 
(driver cell in the following), in which polarization along one of the 
diagonals is externally enforced, also the electrons in the driven cell
will align along the same diagonal, thereby minimizing the total electrostatic
energy. It is therefore possible to propagate the polarization state along a
chain of cells and it has been shown \cite{LentJAP} that
all combinatorial logic functions can be performed by properly designed 
two-dimensional arrays of such cells.

Various implementations of QCA cells have been proposed so far, based 
on metal islands 
\cite{OrlovAPL,OrlovSCI}, on quantum dots obtained in semiconductor 
heterostructures\cite{SmithSCI} or on nanostructured silicon islands
\cite{Prins}. All of these implementations share the same problem: an extreme
sensitivity to fabrication tolerances and the associated need for careful 
adjustment of each single cell. Such a sensitivity is the direct consequence 
of the smallness of the electrostatic interaction between nearby cells and 
therefore of the energy splitting between the configurations corresponding
to the two logic states. While for the purpose of large-scale integration 
new approaches are needed, such as, possibly, the resort to implementations
on the molecular scale, experiments for the assessment of the basic principle
of operation are being performed by carefully 
tuning the voltages applied to adjustment electrodes built in each cell.
The understanding that could so far be gathered from the existing literature
was that strongly bistable and effective QCA operation was possible only in
two regimes: either for cells containing just two electrons\cite{LentJAP} 
(and this would be the case for semiconductor quantum dots) or for cells 
containing two excess electrons on top of a very large total number of 
electrons, and operating in the classical Coulomb blockade limit\cite{LentMet}.

Based on the generally used expression for cell polarization $P$ given in
Ref.\cite{LentAPL} 
\begin{equation}
\label{pollent}
P={{\rho_1+\rho_3-\rho_2-\rho_4}\over{\rho_1+\rho_2+\rho_3+\rho_4}},
\end{equation}
operation was disrupted as soon as
the number of electrons $n$ per cell was other than two ($\rho_i$ is the 
charge in dot $i$, and dots are numbered clockwise). 
For $n > 2$ the maximum 
polarization 
reached by the 
driven cell decreases, due to the fact that, while the denominator of 
Eq.(\ref{pollent}) is $nq$, where $q$ is the electron charge, the 
numerator at most reaches a value of 2$q$. Indeed, a configuration with an 
excess of more than two electrons along one of the diagonals is not 
energetically favored 
for any reasonable arrangement   of neighboring cells. 

We observe that each cell is globally neutral,
because electron charges are compensated for by ionized donors and by the 
positive charge induced on the electrodes defining the quantum dots.
Such neutralization takes place over a certain region of space, with a 
finite extension. Therefore, even though the global monopole component
of the electric field is zero, some effects proportional to the total 
number of electrons contained in the cell exist, but they are
 much weaker than 
those of the uncompensated ``dipole'' component associated with the 
asymmetry between the two diagonals, at least for most configurations of 
practical interest.
This has lead us to proposing a somewhat different expression for cell
polarization, in which the denominator is always 2$q$, independent of total
cell occupancy: 
\begin{equation}
\label{polnos}
P={{\rho_1+\rho_3-\rho_2-\rho_4}\over{2 q}}.
\end{equation}
We argue that Eq.(\ref{polnos}) provides a more realistic representation of 
the action of a cell on its neighbors than Eq.(\ref{pollent}).

If the positive neutralizing charges were in the very same plane as that
of the cell, and localized in each dot in an amount corresponding to $nq/4$, 
as in 
Ref.~\cite{LentJAP}, our statement that only the difference between the
numbers of electrons along the two diagonals matters would be rigorous, 
because the net charge in each dot is the same as in the case of a 2-electron 
cell, in any realistic case.

The situation 
changes somewhat if the neutralizing charge is not located in the same plane 
as that of the cell electrons and/or is not equally distributed among the
dots. In order to show that, in practical operating conditions, 
Eq.(\ref{polnos}) still provides the best description of the polarizing action 
of a cell, we have studied two specific limiting cases: 
(a) neutralization
by means of four $nq/4$ charges located in correspondence with the dots, but on
a plane placed at an arbitrary distance $d$ from the cell;
(b) neutralization by means of image charges located on a plane 
at an arbitrary distance $h$ from that of the cell (such as in the 
case of Dirichlet surface boundary conditions at a distance $h/2$). 

We have first considered a driver cell with a variable number of electrons, 
coupled to a driven cell with just two electrons, and investigated the 
polarization of the latter cell as a function of the polarization of the 
former (defined according to Eq.(\ref{polnos})). For simplicity, we have 
assumed classical point-like charges in the driver cell, while a full
quantum mechanical solution has been performed for the driven cell,
by means of the Configuration-Interaction (CI) method\cite{papsym}. The 
CI technique is based on expanding the many-electron
wave function into a linear combination of Slater determinants built starting
from a single-electron basis. The coefficients of this linear combination
are the unknowns of the problem and can be determined by solving an algebraic 
eigenvalue problem\cite{papsym}, with a dimension corresponding to the
number of Slater determinants that are taken into consideration. We assumed,
for the driven cell, a confinement potential generated in a GaAs/AlGaAs 
heterostructure at a depth of 70~nm by a metal gate with four 90~nm holes
with centers located at the vertices of a 110~nm square, considering
an applied voltage of $-0.5$~V. The distance $D$ between cell centers is
300~nm.

\begin{figure}
\epsfxsize=80mm
\epsffile{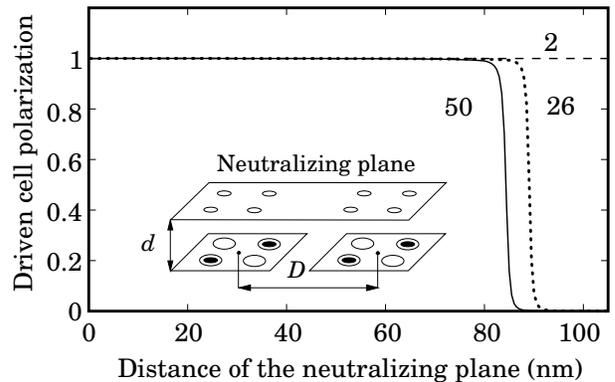}
\caption{Polarization of the driven cell as a function of the distance
$d$ between
the cell and the neutralizing charge plane, for 2, 26 and 50 electrons and a
0.7 polarization of the driver cell.}
\label{eins}
\end{figure}

Let us first examine case (a): in Fig.~\ref{eins} 
we report the polarization of the driven
cell, in response to a 0.7 polarization of the driver cell, as a function
of $d$ for 2, 26 and 50 electrons in the driver cell. 
If there is a total of just two electrons, the driven cell is always fully
polarized, independently of the distance at which the neutralizing charges
are located. When the number of electrons becomes larger, the polarization of
the driven cell is unaffected, as long as the neutralizing charges are
within a reasonable distance from the driver cell; above a certain threshold
value for $d$ (depending on $n$) the locally uncompensated repulsive action
of the electrons in the driver cell prevails and forces the electrons of the
driven cell into the two rightmost dots, thus yielding zero polarization.

\begin{figure}
\epsfxsize=80mm
\epsffile{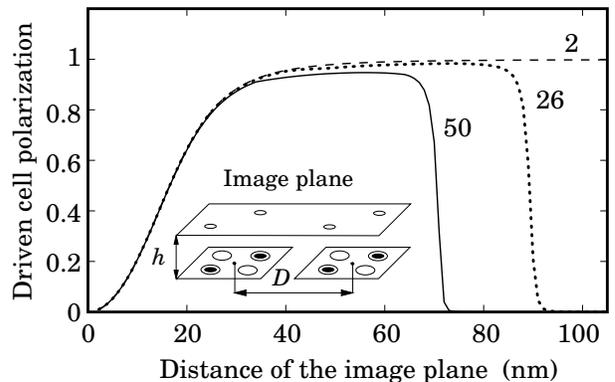}
\caption{Polarization of the driven cell as a function of the distance $h$
between the cell and the image charge plane for 2, 26 and 50 electrons and a
0.7 polarization of the driver cell.}
\label{zwei}
\end{figure}

As far as case (b) is concerned,
in Fig.~\ref{zwei} results are shown 
for 2, 26 and 50 electrons, for the previously described
operating conditions.
For large enough values of $h$, the polarization of the driven cell drops down
to zero, because of the repulsive action of the locally uncompensated charge.
In addition, the polarization decreases (in the same fashion, regardless of the
number of electrons) for decreasing $h$: this is easily understood considering
that the image charges do screen the action of the driver
cell and such screening becomes more effective as the image plane approaches
the cell plane\cite{papsym}.

\begin{figure}
\epsfxsize=80mm
\epsffile{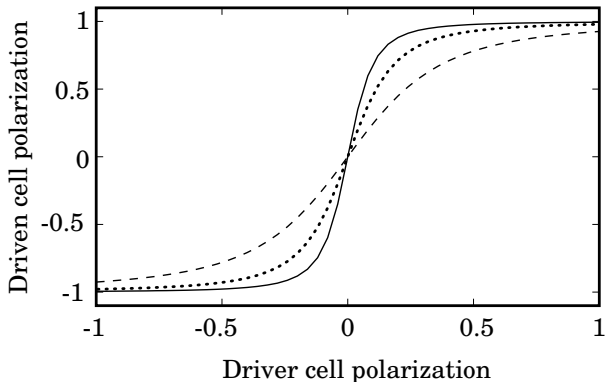}
\caption{Cell-to-cell response function for 2, 26 and 50 electrons in the
driver cell.}
\label{drei}
\end{figure}

In Fig.~\ref{drei} we report the complete cell-to-cell response function,
i.e. the polarization of the driven cell versus that of the driver cell, for
neutralization with image charges at a distance of 70~nm, and for
2, 26 and 50 electrons. Full polarization is reached both for 2 and 26
electrons, with some problem appearing for 50 electrons, that could be
overcome adjusting the geometrical parameters.

Realistic situations are somewhere in between the two cases we have just
discussed, since neutralization is performed by means of charges located both
at the surface (metal gates or surface traps) and in the layers of the
heterostructure. These results confirm the ability of our Eq.(\ref{polnos})
to properly describe the polarizing action of a many-electron driver cell, and
we can move on to the discussion of the response of a many-electron driven cell.

\begin{figure}
\epsfxsize=80mm
\epsffile{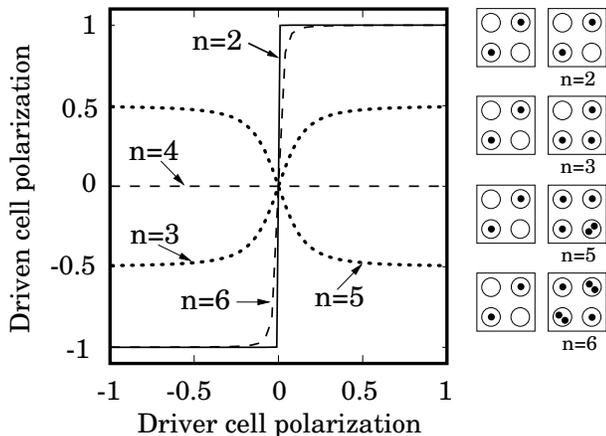}
\caption{Cell-to-cell response function for
a driven cell with 2, 3, 4, 5, and 6 electrons, and sketch of dot occupancy.}
\label{vier}
\end{figure}

For this purpose, we can initially use an intuitive electrostatic model,
in order to gain an immediate understanding of the problem, which will then
be validated with a detailed quantum mechanical calculation.
We consider electrons as classical
particles, interacting via Coulomb repulsion, but with the possibility of
tunneling between dots belonging to the same cell. The driver cell is assumed
to have just two electrons and we examine the response of a many-electron
driven cell. The configurations corresponding to the minimum electrostatic
energy for cells with 3,4,5,6 electrons are shown to the right of 
Fig.~\ref{vier}. It
is apparent that, while for 3 and 5 electrons the maximum polarization is only
one half, and for 4 electrons is zero, for 6 electrons we obtain full
polarization and a behavior that is substantially equivalent to that of
a 2-electron cell.

In order to validate this result, we have performed a quantum mechanical
calculation on cells containing up to 6 electrons, by means of the
CI technique. While for up to 4 electrons a basis
of just 4 single-electron wave functions is adequate for obtaining very
accurate results, for 5 or more electrons a larger basis is in general needed,
because
the presence of two electrons in the same dot leads to a significant deviation
from the single-electron wave functions. An acceptable approximation for
the cases of interest can still be obtained with a total of 8 spin orbitals;
significant improvements in the accuracy require a large increase in the
number of determinants and are beyond the scope of the present
work.

In Fig.~\ref{vier} we report the CI results for the cell-to-cell response
function of driven cells with 2, 3, 4, 5, 6 electrons
for barriers separating 
the dots higher than in the previous cases (this time the voltage applied to 
the gate is -0.7~V): the achieved limiting
polarization values are in exact agreement with the predictions from the
previously presented simple electrostatic model. In addition, we notice
that around the origin the curve for a 2-electron cell is steeper than that
for the 6-electron cell: this is due to the fact that the two ``excess''
(with respect to 4) electrons in the 6-electron cell can be thought of
as ``seeing'' a more shallow confinement potential, resulting from that of
the 2-electron cell plus the electrostatic action of the first four electrons.
Such an effect can be compensated for by raising the potential barriers
separating the dots of each cell.

From the intuitive electrostatic model and from the other results just
described, we can
conclude that QCA cell operation is substantially associated with
the electrons in excess with respect to a multiple of 4: a 6-electron cell
yields full polarization as a 2-electron cell; the same occurs for a
10-electron cell, and, in general, whenever 
the total number of electrons per cell equals $4N+2$,
with $N$ an integer.

This conclusion completes our understanding of the behavior of QCA cells,
filling the gap between the operation with just two
electrons\cite{LentJAP,papsym} and that in the metallic limit, with a
very large number of electrons and two excess charges\cite{LentMet}.
Cells with $4N+2$ electrons are thus suitable for QCA operation, thereby
lowering the technological fabrication requirements; however symmetry
constraints are in no way reduced as a result of the present findings, and
remain the main obstacle preventing the implementation of practicable QCA
logic.
This work has been supported by the ESPRIT project 23362 QUADRANT
(QUAntum Devices foR Advanced Nano-electronic Technology).



\begin{references}
\bibitem{LentAPL}
C. S. Lent, P. D. Tougaw, and W. Porod, Appl. Phys. Lett. {\bf 62}, 714 (1993).
\bibitem{LentJAP}
P. D. Tougaw and C. S. Lent, J. Appl. Phys. {\bf 75}, 1818 (1994).
\bibitem{OrlovAPL}
A. O. Orlov, I. Amlani, G. L. Snider, C. S. Lent, G. H. Bernstein,
Appl. Phys. Lett. {\bf 72}, 2179 (1998).
\bibitem{OrlovSCI}
I. Amlani, A. O. Orlov, G. Toth, G. H. Bernstein, C. S. Lent, G. L. Snider,
Science {\bf 284}, 289 (1999).
\bibitem{SmithSCI}
C. G. Smith, Science {\bf 284}, 274 (1999).
\bibitem{Prins}
C. Single, F. Zhou, H. Heidemeyer, F. E. Prins, D. P. Kern, E. Plies,
J. Vacuum Sci. Tech. {\bf 16}, 3938 (1998).
\bibitem{LentMet}
C. S. Lent and P. D. Tougaw, J. Appl. Phys. {\bf 75}, 4077 (1994).
\bibitem{papsym}
M. Governale, M. Macucci, G. Iannaccone, C. Ungarelli,
J. Martorell, J. Appl. Phys., {\bf 85}, 2962 (1999).
\end{references}
\end{document}